\documentclass[aps,prc,twocolumn,floatfix,nofootinbib,showpacs,superscriptaddress]{revtex4-1}

\usepackage{amsmath,bm}
\usepackage{enumerate}
\usepackage{graphicx}
\usepackage{subfigure}
\usepackage{float}
\usepackage{longtable}
\usepackage{color}

\definecolor{darkgreen}{rgb}{0,0.5,0}
\definecolor{purple}{rgb}{0.5,0,0.5}
\definecolor{nblue}{rgb}{0.0,0.0,0.50}
\definecolor{scarlet}{rgb}{1.0,0.2,0}
\usepackage[colorlinks=true,linkcolor=purple,citecolor=purple,urlcolor=blue]{hyperref}

\begin{document}


\title{Phase diagram and critical endpoint for strongly-interacting quarks}

\author{Si-xue Qin}
\affiliation{Department of Physics and State Key Laboratory of
Nuclear Physics and Technology, Peking University, Beijing 100871,
China}

\author{Lei Chang}
\affiliation{Institute of Applied Physics and Computational
Mathematics, Beijing 100094, China}

\author{Huan Chen}
\affiliation{Institute of High Energy Physics, Chinese Academy of
Sciences, Beijing 100049, China}

\author{Yu-xin Liu}
\affiliation{Department of Physics and State Key Laboratory of
Nuclear Physics and Technology, Peking University, Beijing 100871,
China}
\affiliation{Center of Theoretical Nuclear Physics, National
Laboratory of Heavy Ion Accelerator, Lanzhou 730000, China}

\author{Craig D. Roberts}
\affiliation{Department of Physics and State Key Laboratory of
Nuclear Physics and Technology, Peking University, Beijing 100871,
China}
\affiliation{Physics Division, Argonne National Laboratory, Argonne,
Illinois 60439, USA
}
\date{27 October 2010}

\begin{abstract}
We introduce a method based on the chiral susceptibility, which enables one to draw a phase diagram in the chemical-potential/temperature plane for strongly-interacting quarks whose interactions are described by any reasonable gap equation, even if the diagrammatic content of the quark-gluon vertex is unknown.  We locate a critical endpoint (CEP) at $(\mu^{\rm E},T^{\rm E})\sim (1.0,0.9)T_c$, where $T_c$ is the critical temperature for chiral symmetry restoration at $\mu=0$; and find that a domain of phase coexistence opens at the CEP whose area increases as a confinement length-scale grows.
\end{abstract}


\maketitle


A central goal of the worldwide programme in relativistic heavy ion collisions is to chart the phase diagram of QCD in the plane of nonzero temperature ($T$) and chemical potential ($\mu$).  This will provide fundamental insight into the origin of observable mass and the nature of the early universe.  Two decades of intense speculation have led to an expectation that the phase diagram is complex.  The existence of a critical endpoint in the $(\mu,T)$-plane has been conjectured \cite{Halasz:1998qr}.  In the chiral-limit theory, this marks the end of a line of second-order chiral-symmetry-restoring (and possibly deconfining) transitions, originating on the temperature axis,
and the beginning of a line of first-order transitions.  Such a critical endpoint would have observable consequences \cite{Mohanty:2009vb}.  Hence it is imperative to demonstrate its existence, determine its location and demarcate the subsequent domain of phase coexistence.

Attempts have been made using lattice-QCD.  Such studies rely on Monte Carlo methods but the absence of a probability measure at $\mu\neq 0$ precludes direct computation.  Therefore mathematical devices are necessarily employed in the search for a CEP.  They yield \cite{Fodor:2001pe,Gavai:2004sd,deForcrand:2006ec,Li:2009ju}: $\mu^{\rm E}/T_c = 1.0\,$--$\,1.4$ and $T^{\rm E}/T_c \approx 0.93$, and a signal for a material phase coexistence region \cite{deForcrand:2006ec}.  However, it is not yet certain whether the existence of a CEP survives in simulations with lattice parameters that more closely resemble the physical world \cite{deForcrand:2006pv}.

Models have also been used to search for a CEP.  The Nambu--Jona-Lasinio type yield \cite{Sasaki:2007qh,Costa:2008yh}: $\mu^{\rm E}/T_c \approx 1.7$, $T^{\rm E}/T_c \approx 0.4$; their Polyakov-loop extensions produce \cite{Fu:2007xc,Abuki:2008nm,Schaefer:2008hk,Costa:2009ae}: $\mu^{\rm E}/T_c = 1.5\,$--$\,1.8$, $T^{\rm E}/T_c = 0.3\,$--$\,0.8$; and a chiral quark model gives \cite{Kovacs:2007sy} $(\mu^{\rm E},T^{\rm E})/T_c = (2.0,0.4)$.  On the other hand, a Polyakov-loop-augmented chiral quark model produces \cite{Schaefer:2007pw} $(\mu^{\rm E},T^{\rm E})/T_c = (0.9,0.8)$.
%
%
The former, mutually consistent results for the CEP's location conflict markedly with those obtained from lattice-QCD: $\mu^{\rm E}/T_c$ is significantly larger and $T^{\rm E}/T_c$, much smaller.  If they are nevertheless correct, then finding the CEP in experiment will be difficult because modern colliders are restricted to exploration of the small-$\mu$ domain.
Given this observation, it is unsurprising that an analysis of flow data from the relativistic heavy ion collider leads to the estimate \cite{Lacey:2007na}: $\mu^{\rm E}/T_c \gtrsim 1.0$ and $T^{\rm E}/T_c \lesssim 1.0$.

The Dyson-Schwinger equations (DSEs) provide a nonperturbative approach to studying continuum-QCD \cite{Roberts:1994drRoberts:2000aa} and have been used to prove exact results relating to chiral symmetry \cite{Holl:1998qs,Chang:2008ec,Brodsky:2010xf}.  Simple DSE truncations have been applied to the CEP problem.  A confining zero-width momentum-space interaction, the antithesis of the NJL-model, produces \cite{Blaschke:1997bj} $\mu^{\rm E}/T_c =0$, $T^{\rm E}/T_c = 1$; and a separable-interaction \cite{He:2008yr}: $\mu^{\rm E}/T_c =1.09$, $T^{\rm E}/T_c = 0.78$.  However, neither study described a region of coexisting phases.  Notwithstanding that, in this chain of remarks about the model results, there is a hint that the length-scale characterising confinement in the quark-antiquark interaction markedly influences the location of the CEP.

Herein we employ the DSEs to produce a phase diagram for strongly-interacting quarks, to locate a CEP and demarcate the coexistence region.  The basic tools are the chiral susceptibility and the gap equation.  In QCD its kernel is defined by a contraction of the dressed-gluon-propagator and -quark-gluon-vertex.
For the former we use a form that can interpolate between models of the non-confining NJL-type and the confining interactions used in efficacious DSE studies of hadron observables \cite{Maris:2003vkRoberts:2007jh}, whilst always providing a super-renormalisable interaction.
For the latter, we use either the rainbow-truncation; i.e., the leading-order term in a symmetry-preserving scheme \cite{Bender:1996bb}, or a dressed-vertex \emph{Ansatz}.  The capacity to draw the phase diagram derived from an arbitrary dressed-vertex is essentially new.

At $T\neq 0 \neq \mu$, the gap equation is ($\tilde\omega_n= \omega_n + i \mu$)
\begin{eqnarray}
\label{eq:gap1}
S(\vec{p},\tilde\omega_n)^{-1} &=&  i\vec{\gamma}\cdot\vec{p}
+ i\gamma_4 \tilde \omega_{n} + m
  + \Sigma(\vec{p}, \tilde\omega_{n}) \, ,\\
\nonumber
\Sigma(\vec{p},\tilde\omega_n) &=& T\sum_{l=-\infty}^\infty \! \int\frac{d^3{q}}{(2\pi)^3}\; {g^{2}} D_{\mu\nu} (\vec{p}-\vec{q}, \Omega_{nl}; T, \mu)\\
& & \times \frac{\lambda^a}{2} {\gamma_{\mu}} S(\vec{q},
\tilde\omega_{l}) \frac{\lambda^a}{2}
\Gamma_{\nu} (\vec{q}, \tilde\omega_{l},\vec{p},\tilde\omega_{n})\, ,
\label{eq:gap2}
\end{eqnarray}
where: $\omega_n=(2n+1)\pi T$ is the fermion Matsubara frequency; $\Omega_{nl} = \omega_{n} - \omega_{l}$; $D_{\mu\nu}$ is the dressed-gluon propagator; and $\Gamma_{\nu}$ is the dressed-quark-gluon vertex.  (As we employ an ultraviolet-finite model, renormalisation is unnecessary and $m=0$ in Eq.\,(\ref{eq:gap1}) defines the chiral limit.)

The gap equation's solution can be expressed as
\begin{eqnarray}\label{eq:qdirac}
\nonumber
S(\vec{p},\tilde\omega_n)^{-1} & = & i\vec{\gamma} \cdot \vec{p}\, A(\vec{p}\,^2, \tilde\omega_{n}^2) \\
&& + i\gamma_{4} \tilde\omega_{n} C(\vec{p}\,^2, \tilde\omega_{n}^2) + B(\vec{p}\,^2, \tilde\omega_{n}^2) \, ,
\end{eqnarray}
with, e.g., $B(\vec{p}\,^2, \tilde\omega_{n}^2)^\ast = B(\vec{p}\,^2, \tilde\omega_{-n-1}^2)$.
The dressed-gluon propagator has the form
\begin{equation}
g^2 D_{\mu\nu}(\vec{k}, \Omega_{nl}) = P_{\mu\nu}^{T} D_{T}(\vec{k}\,^2, \Omega_{nl}^2) + P_{\mu\nu}^{L} D_{L}(\vec{k}\,^2, \Omega_{nl}^2)\,,
\end{equation}
where $P_{\mu\nu}^{T,L}$ are, respectively, transverse and longitudinal projection operators.  Whilst for $T\neq 0 \neq \mu$ it is generally true that $D_T \neq D_L$, there are indications \cite{Cucchieri:2007ta} that for $T<0.2\,$GeV, the domain with which we are concerned, it is a good approximation to treat $D_T = D_L=:D_0$.  For the in-vacuum interaction we use a simplified form of that in Ref.\,\cite{Maris:1997tm}; viz., with $\kappa = \vec{k}\,^2 + \Omega_{nl}^2$,
\begin{equation}
\label{IRGs}
D_0(\kappa) = D \frac{4\pi^2}{\sigma^6} \kappa \, {\rm e}^{-\kappa/\sigma^2}.
\end{equation}
The parameters in Eq.\,(\ref{IRGs}) are $D$ and $\sigma$ but they are not independent: a change in $D$ can be compensated by an alteration of $\sigma$ \cite{Maris:2003vkRoberts:2007jh}.  For $\sigma\in[0.3,0.5]\,$GeV, using Eq.\,(\ref{rainbowV}) below, ground-state pseudoscalar and vector-meson observables are roughly constant if $\sigma D  = (0.8 \, {\rm GeV})^3$. 
We usually use $\sigma=0.5\,$GeV.  NB.\ Eq.\,(\ref{IRGs}) is used for illustrative simplicity, not out of necessity.  The status of propagator and vertex studies can be tracked from Ref.\,\cite{Roberts:2007ji}.

The gap equation is complete once the vertex is specified.  For the meson spectrum it is now possible to use any reasonable \emph{Ansatz} \cite{Chang:2009zb}.  Herein we compare results obtained using the rainbow-truncation:
\begin{equation}
\label{rainbowV}
\Gamma_\nu(\vec{q}, \tilde\omega_{l},\vec{p},\tilde\omega_{n}) = \gamma_\nu\,,
\end{equation}
the leading term in a symmetry-preserving scheme \cite{Bender:1996bb}, with those produced by the Ball-Chiu \emph{Ansatz} \cite{Ball:1980ay,Maris:2000ig}:
\begin{eqnarray}
\nonumber
i\Gamma_{\mu}(\vec{q}, \tilde\omega_{l},\vec{p},\tilde\omega_{n}) & = & {i\gamma_{\mu}^{T}}{\Sigma_{A}} + {i\gamma_{\mu}^{L}} {\Sigma_{C}}   \\
\nonumber
& &
+ (\tilde{p}_n+\tilde{q}_l)_{\mu} \bigg[
\frac{i}{2}\gamma^{T}_\alpha (\tilde{p}_n + \tilde{q}_l)_\alpha {\Delta_{A}} \\
& & +
\frac{i}{2}\gamma^{L}_\alpha (\tilde{p}_n+\tilde{q}_l)_\alpha {\Delta_{C}}
+ {\Delta_{B}} \bigg], \label{bcvtx}
\end{eqnarray}
with ($F=A,B,C$)
\begin{eqnarray}
\Sigma_F(\vec{q}\,^2,\omega_l^2,\vec{p}\,^2,\omega_n^2) &= &\frac{1}{2}
[F(\vec{q}\,^2,\omega_l^2)+F(\vec{p}\,^2,\omega_n^2)],\\
\Delta_F(\vec{q}\,^2,\omega_l,\vec{p}\,^2\omega_n) &=&
\frac{F(\vec{q}\,^2,\omega_l^2)-F(\vec{p}\,^2,\omega_n^2)}
{\tilde{q}_l^2-\tilde{p}_n^2}, \label{eq:BC_delta}
\end{eqnarray}
where, defining $u=(0,0,0,1)$, $\gamma_\mu^T=\gamma_\mu-u_\mu\gamma_\alpha u_\alpha$, $\gamma_\mu^L=u_\mu\gamma_\alpha u_\alpha$; and $\tilde{p}_n=(\vec{p},\omega_n+i\mu)$, $\tilde{q}_l=(\vec{q},\omega_l+i\mu)$.
The comparison is natural because vertices of the type in Eq.\,(\ref{rainbowV}) are widely used in studies of hadron observables \cite{Maris:2003vkRoberts:2007jh,Roberts:2007ji}, and the Ball-Chiu \emph{Ansatz} provides a semi-quantitatively accurate representation of lattice-QCD results for important terms in $\Gamma_\mu$ at $T=0=\mu$ \cite{Bhagwat:2004kj}.

\begin{figure}[t]
\centering
\includegraphics[width=8.0cm]{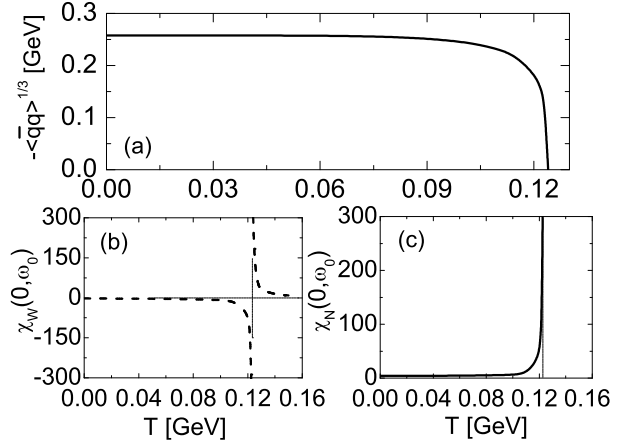}
\caption{Panel~(a) -- Temperature dependence of the chiral-symmetry order parameter in Eq.\,(\protect\ref{qbq}).  Chiral susceptibility computed in the Wigner phase -- Panel~(b), and in the Nambu phase -- Panel~(c).  In all panels: the Ball-Chiu vertex was used, Eq.\,(\protect\ref{bcvtx}); and $\mu=0$, $D=0.5\,$GeV$^2$, $m=0$.
\label{fig:Teffect-BC}}
\end{figure}

We have solved the gap equation formulated above and in Fig.\,\ref{fig:Teffect-BC} depict the $T$-dependence of a chiral susceptibility \cite{Holl:1998qs} and a chiral-symmetry order parameter \cite{Brodsky:2010xf}
\begin{eqnarray}
\chi(0,\omega_0) & = &\frac{\partial}{\partial m} B(\vec{0},\omega_0^2)\,,\\
\label{qbq}
-\langle \bar q q \rangle^0 &= &N_c T \sum^{\infty}_{n= -\infty}\!\!\! {\rm tr}_{\rm D} \int \!
\frac{d^3 p}{(2\pi)^3} \, S_{m=0}(\vec{p},\omega_n) \, .
\end{eqnarray}
For $T<T^{\rm E}$ the behaviour of the order parameter is typical of models without long-range correlations in the gap equation's kernel \cite{Holl:1998qs}.  Namely, initially slow evolution from its $T=0$ value: $\langle \bar q q \rangle^0=(-0.258\,{\rm GeV})^3$, which signals chiral symmetry realised in the Nambu mode; i.e., dynamically broken chiral symmetry, and this followed by a mean-field transition to a phase with chiral symmetry restored; i.e., realised in the Wigner mode.  

The lower panels in Fig.\,\ref{fig:Teffect-BC} show the chiral susceptibility of the Wigner and Nambu phases, which correspond to gap equation solutions that are, respectively, within the domain of attraction of the $B=0$ or $B\neq0$ solution \cite{Chang:2006bm}.  A phase is unstable in response to fluctuations if the susceptibility is negative, but stable and realisable otherwise.  With $\mu=0$, one sees the Nambu phase completely replaced by the Wigner phase at $T=124\,$MeV.

\begin{figure}[t]
\centering
\includegraphics[width=8.0cm]{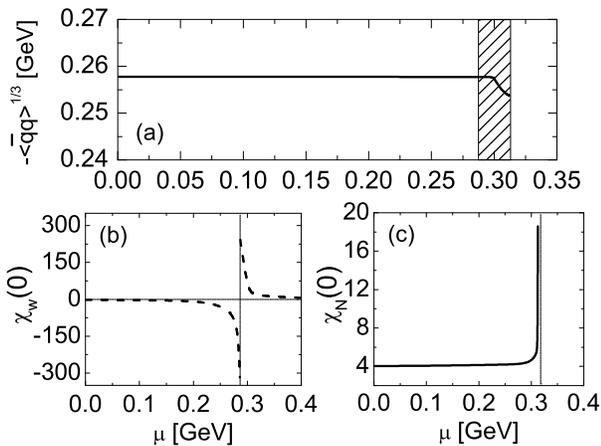}
\caption{Analogue of Fig.\,\ref{fig:Teffect-BC}, displaying evolution with chemical potential at $T=0$.
\label{fig:mueffect-BC}}
\end{figure}

This should be contrasted with the behaviour in Fig.\,\ref{fig:mueffect-BC}.  At $T=0$ the order parameter remains constant with increasing $\mu$ until $\mu_a = 0.30\,$GeV, which is the upper bound on the domain of analyticity for our gap equation's kernel \cite{Chen:2008zr}.  On a small domain beyond this; viz., $\mu\in (\mu_a,\mu_c^N)$, with $\mu_c^N=0.314\,$GeV, the order parameter diminishes smoothly, an effect that may be denominated a partial restoration of chiral symmetry.  For $\mu>\mu_c^N$ the order parameter vanishes so that chiral symmetry is completely restored via a first-order transition.

The lower panels of Fig.\,\ref{fig:mueffect-BC} provide additional information.  For $\mu<\mu_c^W$, with $\mu_c^W = 0.286\,$GeV, the Wigner phase is unstable.  However, that changes at $\mu_c^W$, when the chiral susceptibility in the Wigner phase switches sign and thereafter, on the domain $\mu_c^W < \mu < \mu_c^N$, both the Wigner- and Nambu-phase susceptibilities are positive.  This is the domain of phase coexistence, with a metastable Wigner phase for $\mu_c^W<\mu<\mu_a$ and a metastable Nambu phase for $\mu_a<\mu<\mu_c^N$.  The pressure of the phases is equal at $\mu=\mu_a$; and the Nambu phase is completely displaced by the Wigner phase for $\mu>\mu_c^N$.
Notably, with an \emph{Ansatz} for the dressed-quark gluon vertex, the diagrammatic content of the gap equation's kernel is generally unknown.  However, owing to the insights provided in Ref.\,\cite{Zhao:2008zzi}, one can draw these conclusions despite being unable to calculate an explicit expression for the thermodynamic pressure.

At the onset of the coexistence domain we expect pockets of deconfined, chirally symmetric quark matter to appear in the confining Nambu medium.  Their number and average volume will increase with $\mu$.  The opposite situation occurs at the termination of the domain; i.e., it is the Nambu phase which exists only in pockets.  For $\mu\in (\mu_a,\mu_c^N)$, which is the domain of Nambu-phase metastability, the properties of observed hadrons will be affected by the partial restoration of chiral symmetry.

\begin{figure}[t]
\centering
\includegraphics[width=7.0cm]{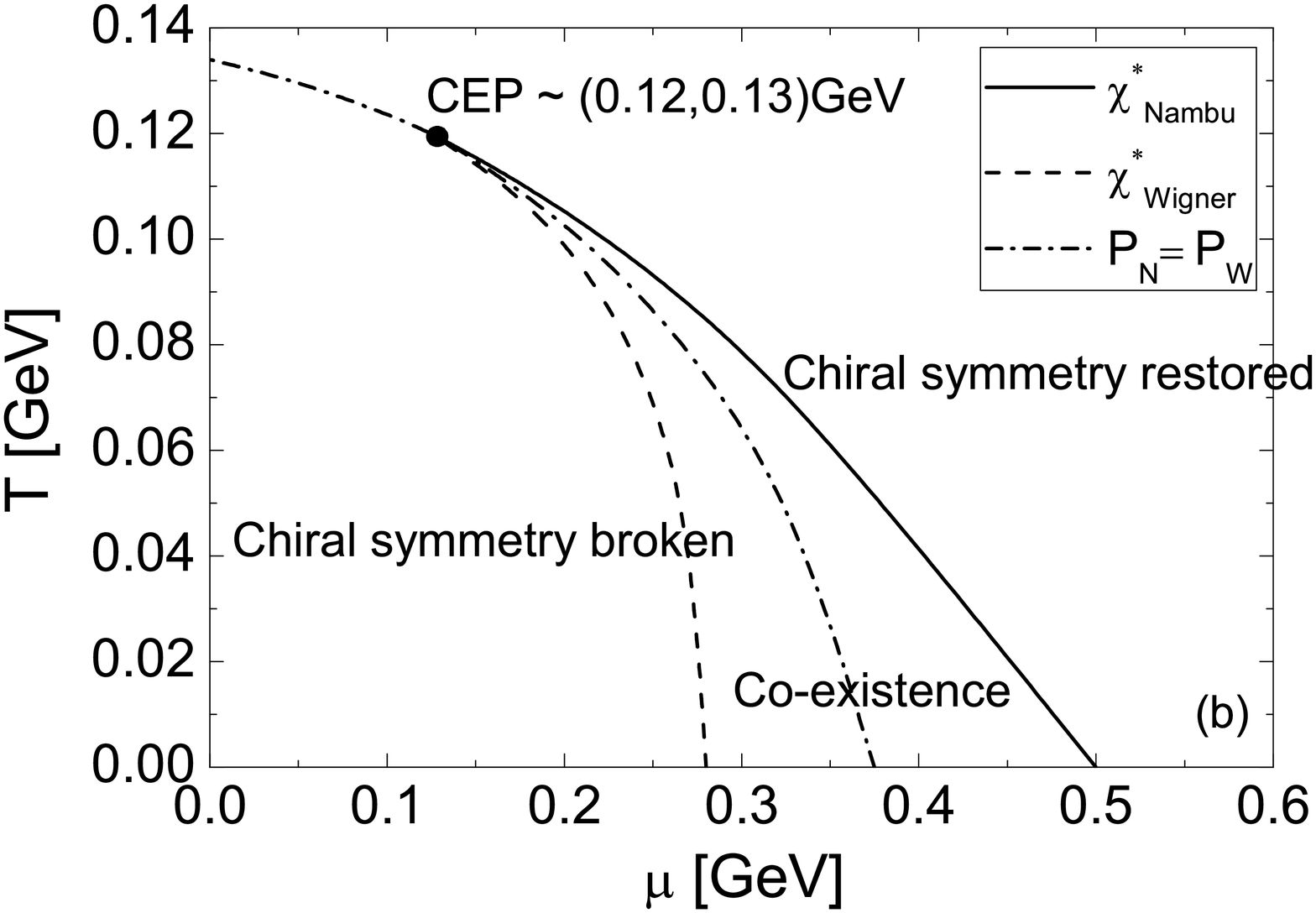}
\includegraphics[width=7.0cm]{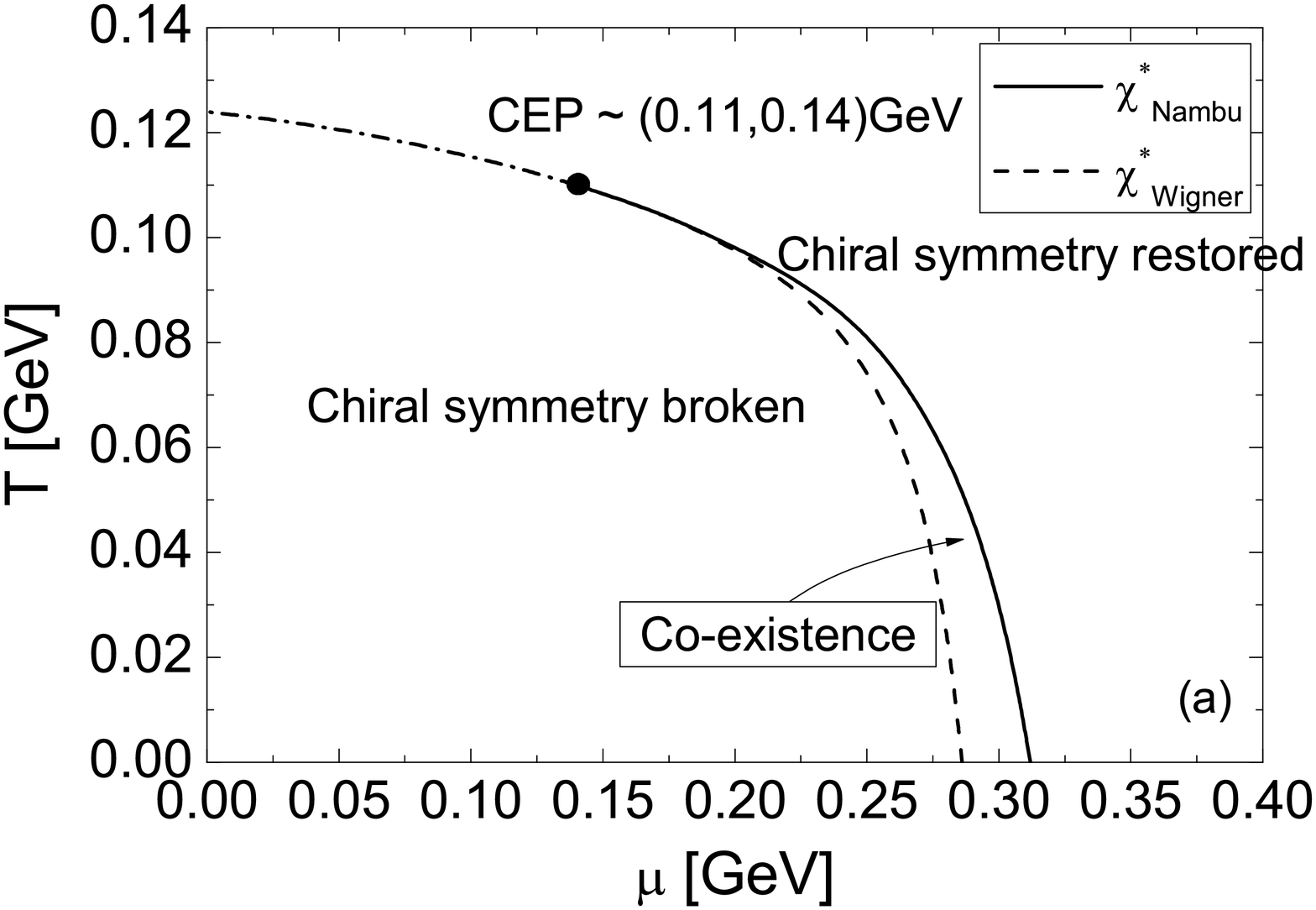}
\caption{Chiral-limit phase diagram in the tem\-pe\-ra\-ture/che\-mi\-cal-po\-ten\-tial plane for strongly-interacting quarks.  The critical endpoint (CEP) is marked explicitly.
\emph{Upper panel} -- rainbow vertex, Eq.\,(\ref{rainbowV}), with $D=1.0\,$GeV$^2$.
%
\emph{Lower panel} -- Ball-Chiu vertex, Eq.\,(\ref{bcvtx}), with $D=0.5\,$GeV$^2$.
%
\label{fig:CEP}}
\end{figure}

We performed computations at many $(\mu,T)$-values and therefrom drew the phase diagrams in Fig.\,\ref{fig:CEP}.  The upper panel was obtained with the rainbow truncation, Eq.\,(\ref{rainbowV}), in which case the diagrammatic content of the gap equation's kernel is known and one can thus compute the dressed-quark component of the pressure.  It is given by the auxiliary-field effective action evaluated at its extrema \cite{Roberts:1994drRoberts:2000aa}.  Within the domain $(\mu<\mu^{\rm E},T>T^{\rm E})$ the system exhibits a mean-field transition, which is signalled both by: equality of the Wigner- and Nambu-phase pressures; and coincident singularities in the Wigner- and Nambu-phase chiral susceptibilities.  The curve tracking the singularity location in the Wigner and Nambu susceptibilities bifurcates at the CEP, with a domain of phase coexistence opening.  Naturally, the curve of equal thermodynamic pressure lies within this domain.

The lower panel in Fig.\,\ref{fig:CEP} was obtained using the dressed vertex, Eq.\,(\ref{bcvtx}).  Its features are similar to those displayed in the upper panel but the domain of phase coexistence is smaller.
Here, as mentioned above, one cannot derive an expression for the dressed-quark pressure.  Thus it is only with an appreciation of the information contained in the chiral susceptibilities that it is possible to draw a phase diagram at all.

It has long been conjectured that confinement is expressed in the analytic structure of the dressed-quark propagator \cite{Roberts:1994drRoberts:2000aa,Maris:2003vkRoberts:2007jh,Roberts:2007ji}.  Measured in this way, it is significant that the models we have considered are members of a class in which chiral symmetry restoration is accompanied by a coincident dressed-quark deconfinement transition in the chiral limit.


In Table~\ref{tab:CEP} we illustrate the response of the CEP's location to changing the vertex or the parameters.  Defining a confinement length-scale $r_\sigma=1/\sigma$, it is apparent that the CEP rotates toward the temperature axis as $r_\sigma$ is increased.  The extreme case is $r_\sigma = \infty$, which was computed in Ref.\,\cite{Blaschke:1997bj} and reported above: ($\mu_{\text{E}}^{},T_{\text{E}}^{})/T_c = (0,1)$.  Models of the NJL-type, as they have been used in the current context, represent the opposite limiting case: they are expressed via a gap equation in which the confinement length-scale vanishes.  From this perspective, it is unsurprising that they produce a CEP whose angular separation from the $\mu$-axis is significantly smaller.

\begin{table}[t]
\begin{center}
\caption{Parameter- and vertex-dependence of the critical endpoint and coexistence region.  $\Delta_{\rm C}$ is the width of the coexistence region on the $\mu$-axis.  Each parameter set produces similar results for the in-vacuum values of the so-called vacuum quark condensate $\langle \bar q q\rangle^0$ and the pion's leptonic decay constant $f_\pi$ \protect\cite{Chang:2009zb}.  Lattice-QCD suggests ($\mu_{\text{E}}^{},T_{\text{E}}^{}$)$/T_c = (1.0\,$--$\,1.4,0.93)$.  (Dimensioned quantities in GeV.)
\label{tab:CEP}}
\begin{tabular}{|c|c|c|c|c|c|c|}
\hline
\multicolumn{3}{|c|}{model} & \multicolumn{4}{|c|}{result} \\
\hline
%
%
vertex & $D^{1/2}$ & $\sigma$ & $T_c$ & $\Delta_{\rm C}$ &($\mu_{\text{E}}^{}$,$T_{\text{E}}^{}$)$/T_c$ & ${\mu_{\text{E}}^{}}/{T_{\text{E}}^{}}$
\\
\hline
BC   & 0.71 & 0.50 & 0.124 & 0.026 &($1.13,0.89$) & $1.27$                    \\
BC   & 0.71 & 0.45 & 0.142 & 0.075 & ($0.85,0.88$) & $0.96$                    \\
Bare & 1.00 & 0.50 & 0.133 & 0.220 & ($0.98,0.90$) & $1.08$                    \\
Bare & 1.02 & 0.45 & 0.138 & 0.280 & ($0.80,0.88$) & $0.90$                    \\
\hline
\end{tabular}
\end{center}
\end{table}

We described a method, based on the chiral susceptibility, which enables one to draw a phase diagram in the chemical-potential/temperature plane for quarks whose interactions are described by any sensibly-constructed gap equation.  Thus, in attempting to chart the phase structure of QCD using the methods of continuum quantum field theory, one is no longer restricted to the simplest class of mean-field kernels: sophisticated quark-gluon vertices can be used.
The method is general and potentially useful in all branches of physics that explore the properties of dense fermionic systems.

A class of models that successfully describes in-vacuum properties of $\pi$- and $\rho$-mesons, exhibits a critical endpoint (CEP) in the neighbourhood $(\mu^{\rm E},T^{\rm E})\sim (1.0,0.9)T_c$.  The CEP's angular separation from the temperature axis is a measure of the confinement length-scale: the separation decreases as the confinement length-scale increases.  Furthermore, a domain of phase coexistence opens at the CEP.  It's size depends on the structure of the gap equation's kernel but, other aspects being equal, it increases in area as the confinement length-scale increases.  Our results suggest that illumination of the CEP and its consequences is within the reach of modern colliders.


%
Work supported by:
National Natural Science Foundation of China, under contract nos.~10425521, 10705002 and 10935001;
Major State Basic Research Development Program contract no.~G2007CB815000;
and U.\,S.\ Department of Energy, Office of Nuclear Physics, contract no.~DE-AC02-06CH11357.



\end{document}